\documentclass{article}
\usepackage[]{amsmath,amssymb}
\usepackage{graphics,epsfig}

\begin{document}

\date{}
\title{Numerical determination of the  entanglement entropy for free fields in the cylinder}
\author{Marina Huerta\footnote{e-mail: marina.huerta@cab.cnea.gov.ar} \\
{\sl Centro At\'omico Bariloche,
8400-S.C. de Bariloche, R\'{\i}o Negro, Argentina}}
\maketitle

\begin{abstract}
We calculate numerically 
the logarithmic contribution to the entanglement entropy of a cylindrical region
in three spatial dimensions for both, free scalar and Dirac fields. 
The coefficient is universal and proportional to the type $c$ conformal anomaly in agreement with recent analytical predictions. 
We also calculate the mass corrections to the entanglement entropy for scalar and Dirac fields in a disk. 
These apparently unrelated problems make contact through the dimensional reduction
procedure valid for free fields whereby the entanglement entropy for the cylinder can be calculated as an integral over 
masses of the disk entanglement entropies. Coming from the same numerical evaluation in the lattice, 
each coefficient is cross checked by the other, testing in this way the two results simultaneously.

\end{abstract}

\section{Introduction}
The entanglement entropy, being a measure of the correlations between two subsystems, 
depends on the geometry of the separating surface, becoming for this reason a quantity with a strong geometric character.
This has inspired a holographic interpretation within the AdS-CFT correspondence 
framework \cite{ryu_taka, ryu_taka2}, which provides a purely geometric way of calculating 
the entropy of conformal field theories. There are several results in the literature where the geometrical properties 
for different sets, dimensions and theories are explored
both, from the quantum field theory and holographic approaches. 
Among these, the coefficient $s$ of the logarithmically divergent term in the entanglement entropy $S_{\log}=s\,\log\epsilon$,
where $\epsilon$ is the ultraviolet cut-off, for general conformal field theories in four dimensions has been found to be
proportional to the type $a$ and $c$ conformal anomaly coefficients \cite{solo} 

\begin{equation}
s=\frac{a}{180} \,\chi(\partial V)+\frac{c}{240 \pi}   \int_{\partial V}(k_i^{\mu \nu}k^i_{\nu \mu} -\frac{1}{2} k_i^{\mu \mu}k^i_{\mu \mu})\,.\label{general}
\end{equation}
Here $\chi(\partial V)$ is the Euler number of the surface,  $k^i_{\mu\nu}=-\gamma^\alpha_\mu \gamma^\beta_\nu \partial_\alpha n^i_\beta$ is 
the second fundamental form, $n^\mu_i$ with $i=1,2$ are a pair of unit vectors orthogonal to $\partial V$, and $\gamma_{\mu\nu}=\delta_{\mu\nu}-n^i_\mu n^i_\nu$ is 
the induced metric on the surface. From (\ref{general}), it can be seen that the sphere and the cylinder are sensitive only to one type of anomaly 
coefficient, $a$ or $c$, respectively. This is  \cite{solo}
\begin{equation}
s=\frac{a}{90}  \,,
\end{equation}
for the sphere, while for a cylinder of length $L$ and radius $R$ it is
\begin{equation}
s= \frac{c}{240}\frac{L}{R}\,.
\label{coefs}
\end{equation}
For spherical sets, this result was later recalculated, numerically and analytically \cite{num,dowker,cashue,solo2,dowker1}, using different technics, and recently 
extended to any theory \cite{cashuemye}. Instead, the validity of the result (\ref{coefs}) for the cylinder is more subtle. It has been also studied from 
the holographic point of view in \cite{Myers2} where $s$ is found to be proportional to the $c$ conformal anomaly in four dimensions as in (\ref{coefs}), but it 
cannot be generalized to higher dimensions where other conformally 
invariant terms can be added to (\ref{general}).

In this paper, we study numerically the logarithmic contributions to the entropy for a cylinder in $(3+1)$ dimensions for free massless 
scalar and Dirac fields. The method we use consists first, in dimensionally reducing the problem of the three dimensional cylinder to the one of an infinite set of massive fields 
living in a two dimensional spatial disk. The entanglement entropy is then calculated numerically by the real time approach \cite{review} where the 
reduced density matrix is written in terms of free correlators. This is based, on one hand, on the method developed by Sredniki 
for calculating the entropy for scalars in spherical sets \cite{sred} where the field is discretized in the radial direction in polar coordinates, and 
on the other, on the work by Peschel, 
whereby the reduced density matrix can be written in terms of correlators for both, bosonic and fermionic, solvable lattice systems \cite{peschel}.  
Thus, for the numerical evaluation of the entropy of a Dirac field in the disk, we extend the method described in \cite{sred}, originally applicable only to scalar fields.  

Then, from the expansion of the entropy 
in powers of $(mR)$ valid in the large $mR$ limit, we find the coefficient $s$ for the cylinder (\ref{coefs}).
This is directly related to the coefficient of 
the term $(mR)^{-1}$. In the same expansion, we also identify the coefficient of the linear term $mR$. This last, has been recently calculated analytically 
in \cite{wilczek}, where the same type of expansion is considered for general smooth geometries and massive scalar theories. From the holographic point of view, 
new contributions to the entanglement entropy were also found 
when considering theories deformed by a relevant operator \cite{Myers3}, consistently to that reported in \cite{wilczek}. 

We find, for the cylinder, within a porcentual error $\sim 0.15 \%$ for scalars and $\sim 1\%$ for Dirac fields,
\begin{eqnarray}
s_{s}&=&\frac{L}{240R}\,, \nonumber \\
s_{f}&=&\frac{6L}{240R}\,, \label{cyl}
\end{eqnarray}
and for the disk, within a porcentual error $\sim 0.05\%$, 
\begin{equation}
\Delta S_m=-\frac{2 \pi }{12}R~ m\,, \label{disk}
\end{equation}
in agreement  with \cite{solo} and \cite{wilczek}. Coming from the same expansion, 
each coefficient is cross checked by the other testing both at the same time.    
This concordance sums support to the general validity of the method used in \cite{solo}. 
We will discuss this issue in more detail, later in the Discussion. 
  
The paper is organized as follows: in the second Section, we discuss the dimensional reduction procedure and show the problem of massless fields in the cylinder 
can be reduced to the one of massive fields in a disk. In the third Section, we discuss the approach we use for the numerical evaluation of the entanglement entropy
in the disk, for both, scalar and Dirac massive fields. In Section 4 we present our results  and finally, its interpretation in the Discussion.

\section{The cylinder by dimensional reduction}
For free fields, some universal terms in the entanglement entropy in high dimensions can be obtained 
via dimensional reduction technics from results calculated in lower dimensions \cite{review}.
Let us consider a set in three spatial dimensions of the form $V=D\times X$, where $X$ is a line on the first coordinate $x_1$, of length $L$, 
and $D$ is a sphere in two dimensions (disk). 
We are interested in the entropy of $V$ in the limit of large $L$. The direction $x_1$ can be compactified by imposing periodic 
boundary conditions $x_1\equiv x_1+L$, without changing the result of the leading extensive term. For a free massless field we 
Fourier decompose it into the corresponding field modes in the compact direction 
\begin{equation}
\phi_n(x_1,x_2,x_3,t)=\tilde\phi_n(x_2,x_3,t)e^{i\frac{ 2 \pi x_1 n}{L}}\,.
\end{equation}

The problem then reduces to a two dimensional one with an infinite tower of massive fields. 
For example, for a free scalar we obtain 
the tower of fields $\tilde\phi$, 
\begin{equation}
\Box_4 \phi_n=\Box_3 \tilde\phi_n+(\frac{2 \pi n}{L})^2 \tilde\phi_n\,.
\end{equation} 
From the point of view of the non compact $x_{2}, x_{3}$ directions, these fields have masses given by 
\begin{equation}
m_{n}^2=\left(\frac{2\pi}{L}n \right) ^2\,. 
\end{equation}
Summing over the contributions of all the decoupled $2$ dimensional fields we have
\begin{equation}
S(V)=\sum_{n=-{\infty}}^\infty S(D,m_{n})\,.
\end{equation}
In the limit of large $L$ we can convert this sum into integral
\begin{equation}
S(V)= \frac{L}{\pi}\int_0^\infty dm\, S(D,m)\,.\label{ala}
\end{equation}
The universal terms in $S(V)$ will then come from the ones of $S(D)$ 
after integrating over the mass. 
For a Dirac field, the spin multiplicity factor $2^{[(d+1)/2]}$ has to be incorporated. In the present case, it is 

\begin{equation}
S(V)= 2 \frac{L}{\pi}\int_0^\infty dm\, S(D,m)\,\label{ala1}\,.
\end{equation}
Expanding the entanglement entropy $S(D,m)$ in powers of $mR$ for large $mR$ 
\begin{equation}
S(D,m)= c_1 mR +c_0+c_{(-1)}\frac{1}{mR}+...\label{expansion}\,,
\end{equation}
and inserting (\ref{expansion}) in (\ref{ala}) and (\ref{ala1}) for scalar and Dirac fields respectively, we obtain 
for the $(3+1)$  dimensional theory that the logarithmic coefficient $s$ in $S(V)$ is directly related to $c_{(-1)}$ by
\begin{eqnarray}
s_{s}&=& -c^{s}_{(-1)}\frac{L}{\pi R}\,,\\
s_{f}&=&-2 c^{f}_{(-1)}\frac{L}{\pi R} \,.\label{sm}
\end{eqnarray}
On the other hand, for the dimensionally reduced theory, the contribution in the entropy proportional to the mass is given by the term proportional to the coefficient $c_1$. 
This contribution was calculated in 
\cite{wilczek} for massive scalar fields in any dimension for
a waveguide geometry with specified boundary conditions using heat kernel methods. The terms extensive in the area $A_{d-1}$ in even spatial dimensions are given by 
\begin{equation}
 S=\frac{A_{d-1}}{12}\int_{\epsilon^2}^{\infty}\frac{dt}{t}\frac{1}{(4\pi t)^{\frac{d-1}{2}}}e^{-tm^2} \,.\label{heatkernel}
\end{equation}
From (\ref{heatkernel}) it follows,
\begin{equation}
\Delta S_m=\gamma_d~ m^{d-1}~A_{d-1}\,,\\
\end{equation}
where $\gamma_d\equiv(-1)^{(d/2)}[12 ~(2\pi)^{(d-2)/2}(d-1)!!]^{-1}$ for $d$ even.
For Dirac fermions with $2^{[(d+1)/2]}$ components, the same calculation can be done taking into account the extra factor $2^{[(d+1)/2]-1}$ \cite{kabat}, 
which relates the scalar and Dirac $\gamma$ coefficients,
\begin{equation}
\gamma_d^{f}\equiv 2^{[(d+1)/2]-1}\gamma_d^{s}.
\end{equation}
In our case, $d=2$ and 
\begin{equation}
\Delta S_m=c_1 R~ m\,,\,\,\,\,\,\,\,\,\,\,\,\,\,\, c_1=2\pi \gamma_2=-\frac{2 \pi }{12},
\end{equation}
both for Dirac and scalar fields.
 
\section{The disk: Numerical evaluation for massive scalar and Dirac fields}
We evaluate numerically the entanglement entropy for a two dimensional disk. In this section we describe 
the numerical method and models in the lattice for scalar and Dirac massive fields.  
\subsection{Massive scalars in a disk}

Consider the quadratic Hamiltonian for a massive scalar in $(2+1)$ dimensions
\begin{equation}
H=\frac{1}{2}\int dV ((\partial_t \phi)^2+ (\bigtriangledown{\phi})^2 +m^2 \phi^2)\,.
\end{equation}
Following \cite{sred}, we separate variables in polar coordinates and introduce new ones 
\begin{eqnarray}
 \tilde\phi_n(r)&=&\sqrt{\frac{r}{2\pi}}\int d\theta ~e^{i n\theta} \phi(\theta,r)\,,\\
\tilde\pi_n(r)&=&\sqrt{\frac{r}{2\pi}}\int d\theta~ e^{i n\theta} \pi(\theta,r)\,,\\
\end{eqnarray}
such that
\begin{equation}
 [\tilde\phi_n(r),\tilde\pi_{n^{\prime}}(r^{\prime})]=i \delta_{nn^{\prime}}\delta(r-r^{\prime} )\,.
\end{equation}
Then, the Hamiltonian $H=\sum_n H_n$ takes the form
\begin{equation}
 H_n=\frac{1}{2}\int_0^{\infty} dr [\tilde\pi_n^2+r\partial_r(\frac{\tilde\phi_n}{\sqrt{r}})^2+m^2\tilde\phi_n^2+\frac{n^2}{r^2}\tilde\phi_n^2]\,.
\label{hscalar}
\end{equation}

 For a general quadratic discrete Hamiltonian 
$H=\frac{1}{2}\sum \pi _{i}^{2}+\frac{1}{2}\sum_{ij}\phi _{i}K_{ij}\phi
_{j}$, (dropping the $n$ index temporarily), the vacuum (ground state) two point correlators $X_{ij}$ and  $P_{ij}$
\begin{eqnarray}
\left\langle \phi _{i}\phi _{j}\right\rangle &=& X_{ij} \,,\label{atre} \\
\left\langle \pi _{i}\pi _{j}\right\rangle &=&P_{ij} \,, \label{atro}\,\label{need}
\end{eqnarray}
are given by \cite{review}
\begin{eqnarray}
X_{ij} &=&\left\langle \phi _{i}\phi _{j}\right\rangle =\frac{1}{2}(K^{-
\frac{1}{2}})_{ij}\,,  \label{x} \\
P_{ij} &=&\left\langle \pi _{i}\pi _{j}\right\rangle =\frac{1}{2}(K^{\frac{1
}{2}})_{ij}\,, \label{p}
\end{eqnarray}
in terms of the $K$ matrix. Here, $1 \leq i,j\leq N$, where $N$ is the size of the lattice and acts as an infrared regulator.  

Then, the entropy of the disk is given by 
\begin{equation}
S^D\,=S_0+\sum_{n=1}^{\infty} 2 S_n\,,
\label{sscalar}
\end{equation}
with \cite{review}
\begin{equation}
S_n\,= \textrm{tr}\left(( \sqrt{X_n^D P_n^D}+\frac{1}{2})\log (\sqrt{X_n^D P_n^D}+\frac{1}{2})-(\sqrt{X_n^D P_n^D}-\frac{1}{2})\log (\sqrt{X_n^D P_n^D}-\frac{1}{2})\right)\,.  \label{for}
\end{equation}
The superscript $D$ in (\ref{for}) means the indices of the matrices are restricted to the region $D$: $1\leq i,j \leq r$, with $r$ the radius of the disk.

In this case, after discretization we find  the matrix $K_n$ corresponding to (\ref{hscalar}) is
\begin{eqnarray}
 K_n^{11}&=&\frac{3}{2}+n^2+m^2\,,\\
K_n^{ii}&=&2+\frac{n^2}{i^2}+m^2\,,\\
K_n^{i,i+1}&=&-\frac{i+1/2}{\sqrt{i(i+1)}}=K_n^{i+1,i}\,.
\end{eqnarray}

Summarizing, the numerical evaluation of the entropy for massive scalar fields in a disk starts with the calculation of the $(N\times N)$ matrix  $K_n$ 
for a given mass $m$, angular momentum $n$, and infrared regulator $N$ which gives the size 
of the unidimensional lattice. From (\ref{x}) and (\ref{p}) we calculate the two point correlators $X$ and $P$. Then, we reduce them to the disk and calculate the 
contribution $S_n$ (\ref{for}). Finally, the entropy $S^D$ is given by the sum (\ref{sscalar}). 

\subsection{Massive fermions in a disk.}

The Hamiltonian for a massive Dirac field in $(2+1)$ dimensions can be written as 
\begin{equation}
 H=\int dV \psi^{\dagger}(x,y) H^{(1)} \psi(x,y)\,,
\end{equation}
with $H^{(1)}$ the one particle Hamiltonian
\begin{equation}
 H^{(1)}\equiv i \frac{\partial}{\partial_t}=\frac{1}{i}(\alpha_x\partial_x+\alpha_y\partial_y)+\beta m\,.
\end{equation}
Choosing
\begin{eqnarray}
\alpha_x&=&\sigma_1\,,\\
\alpha_y&=&\sigma_2\,,\\
\beta&=&\sigma_3\,,
\end{eqnarray}
with $\sigma_i$ the Pauli matrices, the relations $\{\alpha_i,\alpha_j\}=\{\alpha_i,\beta\}=0$ 
and $\alpha_i^2=\beta^2=1$ are satisfied. In polar coordinates $r,\theta$ the Hamiltonian takes the form

\begin{equation}
 H^{(1)}=\frac{1}{i} \left(\frac{1}{r}h_{\theta}\partial_{\theta}+h_{r}\partial_r \right)+m h_m\,,
\end{equation}
where $h_{\theta}$, $h_r$ and $h_m$ are
\begin{eqnarray}
 h_{\theta}&=& \left( 
\begin{array}{cc}
0 & ie^{i\theta}  \\
-ie^{-i\theta} & 0 \end{array} \right)\,,  \\
 h_{r}&=& \left( 
\begin{array}{cc}
0 & e^{i\theta}  \\
e^{-i\theta} & 0 \end{array} \right) \,,\\
h_m&=&\left( 
\begin{array}{cc}
1 & 0  \\
0 & -1 \end{array} \right) \,.
\end{eqnarray} 
Separating variables, we propose the following two component Dirac spinor
\begin{equation}
 \psi_n=\left(\begin{array}{c}
u_n(r)~ \phi^1_n(\theta)  \\
v_n(r) ~\phi^2_n(\theta)  \end{array} \right) \,,
\end{equation}
where $\phi^1_n=\frac{1}{\sqrt {2\pi}}e^{i\theta(n+\frac{1}{2})}$ and $\phi^2_n=\frac{1}{\sqrt {2\pi}}e^{i\theta(n-\frac{1}{2})}$ 
are the eigenvector components of the angular momentum operator $J=\frac{1}{i}\partial_{\theta}-\frac{1}{2}\sigma_3 $ 
with half integer eigenvalue $n$ 
\begin{equation}
 J\left(\begin{array}{c}
\phi_n^1  \\
\phi_n^2  \end{array}\right)=n\left(\begin{array}{c}
\phi_n^1  \\
\phi_n^2  \end{array}\right)\,.
\end{equation}
Note that $J$ commutes with $H$. 
Then, we can express the Hamiltonian as a sum over $n$ such that 
\begin{equation}
 H= \sum_n H_n=\sum_n \int dr~ r~\alpha^{\dagger}_n(r)H^1_n\alpha_n(r)
\end{equation}
where
\begin{equation}
 H^{(1)}_n=-\frac{1}{r}(i\sigma_1/2+n\sigma_2)-i\sigma_1\partial_r+m\sigma_3\,,
\label{hfer}
\end{equation}
and
\begin{equation}
 \alpha_n(r)=\left(\begin{array}{c}
u_n(r)   \\
v_n(r)  \end{array} \right)\,.
\label{hfer}
\end{equation}

In general, in the discrete free case, the reduced density matrix can be written in terms of a Hermitian operator ${\cal H}$ 
\begin{equation}
 \rho=K e^{-{\cal H}}=Ke^{\sum_V \psi_i^{\dagger}{\hat{\cal H}}_{ij}\psi_j}\,.
\end{equation}
This can be expressed in terms of the correlators $C_{ij}$
\begin{equation}
\langle\psi_i\psi_j^{\dagger}\rangle=C_{ij}\,,
\label{corf}
\end{equation}
via the identification \cite{review}
\begin{equation}
\hat {\cal H}=-\log(C^{-1}-1)\,.
\end{equation}
Here, the $\psi_i$ are fermion operators canonically normalized. Then, the entropy can be written in terms of $C$ as $S= -\text{tr}((1-C)\log(1-C)+C \log C)$.

For a general quadratic case, with discrete Hamiltonian $H=\sum_{ij}M_{ij}\psi_i^{\dagger}\psi_j$, the correlator is directly related to $M_{ij}$ by
\begin{equation}
C=\Theta(-M) \,.
\label{cfe}
\end{equation}
In (\ref{cfe}), the indices run over the complete space. Then, we introduce an infrared regulator $N$ which is the size of the lattice.

Therefore, the entropy $S$ for the disk $D$, is given by a sum over the angular momentum $n$ as
\begin{equation}
 S(D)=\sum_n -\text{tr}((1-C^D_n)\log(1-C^D_n)+C^D_n \log C^D_n)\,.
\label{entd}
\end{equation}
The correlators $C^D_n$ for fixed angular momentum $n$ are restricted to the disk region $D$.

In order to identify the $M$ matrix in the case of the disk, we first introduce the operators $\tilde{\alpha}(r)=r^{(1/2)}\alpha(r)$.
These are normalized such that they satisfy the canonical anticonmutation relations required for the application of (\ref{entd}). 
The discrete Hamiltonian in these variables 
for a $N$ lattice, takes the form
(dropping the angular momentum index temporarily) 
\begin{equation}
 H=\sum_{i,j=1}^NH_{i,j}=\sum_{i,j=1}^N(\tilde u^{*}_i,\tilde v^{*}_i)M_{i,j}\left(\begin{array}{c} \tilde u_j  \\
\tilde v_j  
\end{array} \right)\,,
\end{equation}
where the indices $i,j$ are the discrete variables corresponding to the continuum radial coordinate $r$. $M_{i,j}$, for fixed $i,j$, is a $(2 \times 2)$ matrix such that $H_{i,j}=
M^{1,1}_{i,j}\tilde u_i^{*}\tilde u_j+M^{1,2}_{i,j}\tilde u_i^{*}\tilde v_j+M^{2,1}_{i,j}\tilde v_i^{*}\tilde u_j+M^{2,2}_{i,j}\tilde v_i^{*}\tilde v_j$ with $i,j=1,...,N$. 

Finally, we define a $2N\times2N$ matrix $\tilde{M}_n^{2k+\alpha-2,2l+\beta-2}=M_{k,l}^{\alpha,\beta}$, for $k,l=1,...,N$ and $\alpha,\beta=1,2$. The non zero entries of $\tilde{M}$ 
(for each angular momentum $n$) are
\begin{eqnarray}
\tilde{M}_n^{kk}&=&(-1)^{k+1} m,\\
\tilde{M}_n^{1,2}=i(n+\frac{1}{2}) &,& \tilde{M}_n^{2,1}=-i(n+\frac{1}{2}),\\
\tilde{M}_n^{2k-1,2k}=i~\frac{n}{k} &,& \tilde{M}_n^{2k,2k-1}=-i~\frac{n}{k},\\
\tilde{M}_n^{2k-1,2k+2}=\frac{-i}{2}&,& \tilde{M}_n^{2k,2k-3}=\frac{i}{2},\\
\tilde{M}_n^{2k-1,2k-2}=\frac{i}{2}&,& \tilde{M}_n^{2k,2k+1}=\frac{-i}{2}.
\end{eqnarray}
The $\tilde{M_n}$ matrix satisfies $\tilde{M_n}^{\dagger}=\tilde{M_n}$ and it has real eigenvalues symmetric with respect to the origin. 
There is also a symmetry related to the mapping $n \rightarrow -n$ which can be seen also in the continuum limit.

\begin{figure} [tp]
\leavevmode
\centering
\epsfxsize=8cm
\bigskip
\epsfbox{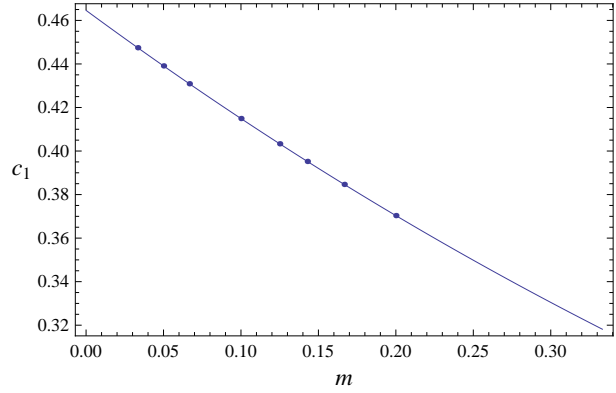}
\caption{Scalar field: The points correspond to the coefficient of the linear term in $r$ in the disk entanglement entropy for different masses. 
The linear coefficient in $m$ in the fit, drawn with a solid line, is $-0.52359\sim -\frac{2\pi}{12}$. This is the value of $c_1$ in (\ref{expansion}). }
\label{figu0}
\end{figure}

\begin{figure} 
\leavevmode
\centering
\epsfxsize=8cm
\bigskip
\epsfbox{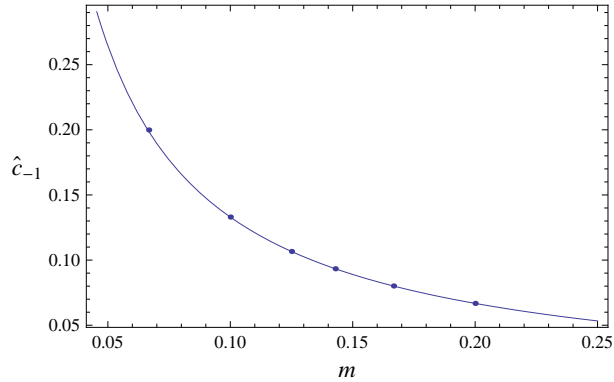}
\caption{Scalar field: $\hat{c}_{-1}(m)=-c_{-1}(m)$. The points correspond to the coefficient of the term $\frac{1}{r}$ in the disk entanglement entropy for different masses. 
The coefficient of the term proportional to $\frac{1}{m}$ in the fit drawn with a solid line is $0.01342 \sim \frac{\pi}{240}$. 
This is the value of $-c_{-1}$ in (\ref{expansion}).}
\label{figu0}
\end{figure}
Summarizing, the numerical evaluation of the entropy for massive Dirac fields in a disk starts with the calculation of the $(2N\times 2N)$ matrix  $\tilde M_n$ 
for a given mass $m$, angular momentum $n$, and $N$, the size 
of the unidimensional lattice. From (\ref{cfe}) we calculate the two point correlator $C$ (\ref{corf}). Then, we reduce it to the disk $C^D_n$ and calculate the 
contribution $S_n$ in (\ref{entd}). Finally, the entropy $S^D$ is given by the sum (\ref{entd}). 

\section{Results}

\begin{figure}[tp]
\leavevmode
\centering
\epsfxsize=8cm
\bigskip
\epsfbox{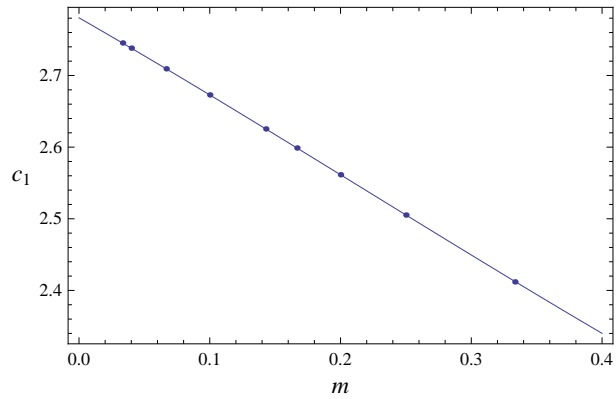}
\caption{Dirac field: The points correspond to the coefficient of the linear term in $r$  in the disk entanglement entropy for different masses. 
The linear coefficient in $m$ in the fit, drawn with a solid line, is $-1.04658 \sim-2\times\frac{2 \pi}{12}$. This is $c_1$ in (\ref{expansion}).
The extra factor of two is due to the fermion doubling in the lattice.}
\label{figu0}
\end{figure}
We calculate the entanglement entropy for a disk in a lattice of $200$ points. We consider regions of radii $r$ going from $50$ to $80$ in lattice units 
for the scalar and from $30$ to $50$ for the fermionic field.
The sum over the angular momentum $n$ is done exactly up to $n_{max}=3000$ and the corrections coming from contributions $n>n_{max}$ are added by 
fitting the exact entropy for some large values of $n$.  
The fit we use is of the form
\begin{equation}
s_n=a_2 \frac{1}{n^2}+b_2 \frac{\log n}{n^2}+a_4 \frac{1}{n^4}+b_4 \frac{\log n}{n^4}+a_6 \frac{1}{n^6}+b_6 \frac{\log n}{n^6}
\end{equation}
since for large $n$ ($n \gg N$) both $C$ and $\sqrt{XP}$ defined in (\ref{cfe}), (\ref{x}) and (\ref{p}) have expansions in even inverse powers of $n$ \cite{sred}. 

This is done for different masses in the range $1/20 < m < 1/2$ such that $m^{-1}>\epsilon$ (we fix $\epsilon=1$) and $m r > 1$.
For each mass, we fit the entropy in terms of the size $r$ of the disk,
\begin{equation}
 S=c_1(m)r+c_0(m)+c_{-1}(m)\frac{1}{r}+...\,.
\end{equation}

Once the coefficients $c_1(m)$ and $c_{-1}(m)$ are identified, we 
expand in powers of $m$,
\begin{eqnarray}
 c_1(m)&=&c_1 m+c^0_1+c^{-1}_1\frac{1}{m}\,,\label{c1m}\\
c_{-1}(m)&=&c^1_{-1} m+c^0_{-1}+c_{-1}\frac{1}{m}\,.\label{c-1m}
\end{eqnarray}
\begin{figure}[tp]
\leavevmode
\centering
\epsfxsize=8cm
\bigskip
\epsfbox{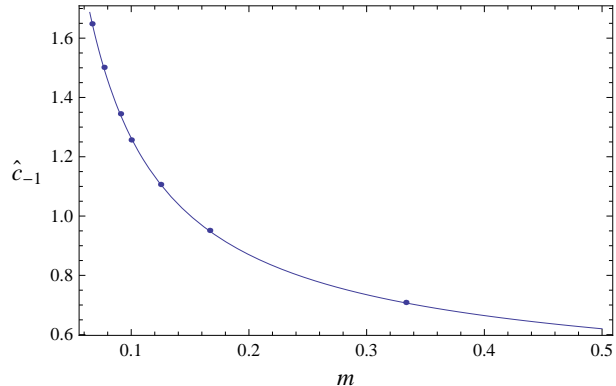}
\caption{Dirac field:  $\hat{c}_{-1}(m)=-c_{-1}(m)$. The points correspond to the coefficient of the term $\frac{1}{r}$ in the disk entanglement entropy for different masses. 
The coefficient of the term proportional to $\frac{1}{m}$ in the fit, drawn with a solid line, is $0.07754 \sim \frac{6\pi}{240}$. 
This is the value of $-c_{-1}$ in (\ref{expansion}). The factors $2$ and $1/2$ due to the fermion doubling 
and the spin multiplicity in the dimensional reduction formula (\ref{sm}) respectively, compensate each other.}
\label{figu0}
\end{figure}
We obtain $c_1$ and $c_{-1}$ of (\ref{expansion}) as the $m$ and $\frac{1}{m}$ coefficients in (\ref{c1m}) and (\ref{c-1m}) respectively. 
For example, in the scalar case, for $m=1/10$ we calculate the entanglement entropy for disks with radii from $60$ to $80$ (in lattice units).
From the best fit, we extract the coefficients of the terms proportional to $r$ and $1/r$. In this example, this gives $ 0.20744 - 0.13303/r + 0.41493 r$. 
Then, $c_1(m=1/10)=0.41493$ and $c_{-1}(m=1/10)=-0.13303$. The values of $c_1(m)$ and $c_{-1}(m)$ are shown in Figures (1),(2) for scalars and Figures (3),(4) 
for the Dirac field.
In order to compare the lattice results with the continuum expectations for fermions, a factor $1/2$ has to be incorporated in the entropy formula due to the 
fermion doubling in the unidimensional radial lattice. 
We obtain
\begin{eqnarray}
c^{s}_{(-1)}&=&-0.01342\sim-\frac{\pi}{240} \,\,,\,\text{within a porcentual error of $0.15$}\,,\label{ana1}\\
c^{f}_{(-1)}&=&-0.07754\sim -\frac{6 \pi}{240}\,\,,\,\text{within a porcentual error of $1.2$} \,,\label{ana2}
\end{eqnarray}
and 
\begin{eqnarray}
c^{s}_1&=&-0.52359\sim -\frac{2 \pi }{12}\,\,\,,\,\text{within a porcentual error of $0.02$}\,,\\ \label{ana3}
c^{f}_1&=&-0.52329\sim-\frac{2 \pi }{12}\,\,\,,\,\text{within a porcentual error of $0.05$}\,,\label{ana3}
\end{eqnarray}
in agreement to the analytical results.

\section{Discussion}

Our numerical results agree with the ones predicted analytically within porcentual errors $\sim 1.2\%$ or less. 
Since both results, the one for the cylinder and 
the one for the disk, come from the identification of different coefficients 
in the same expansion (\ref{expansion}), they cross check each other giving a solid basement to our numerical test. 

On the other hand, by the analytical side within the QFT approach, the general grounds the result for the cylinder
is based on, rely first on the application of the replication method, then on demanding conformal invariance of 
the logarithmic contribution in the entropy derived from the effective action of
the replicated manifold, and finally on calibrating a free parameter by using the holographic 
correspondence \cite{ryu_taka}. 
The replica trick is a standard approach to calculate the entanglement entropy \cite{hlw}. 
In this construction, the trace of the integers powers 
of the reduced density matrix corresponds to the partition function of $n$ copies of the system connected consistently 
through the cuts along the boundary. The replicated manifold is non trivial due to the conical singularities 
of deficit angle $\alpha=2 \pi (n-1)$ placed on the boundary. From this, the entropy associated to the set $V$ can be calculated as
\begin{equation}
S=-\lim_{n \rightarrow 1} \frac{\partial}{\partial n}\textrm{tr}[\rho ^n _V]=-\lim_{n \rightarrow 1}(\frac{\partial}{\partial n}-1)\log Z_n\,.
\label{ee}
\end{equation}
In the above formula we assume that not only the partition function $Z_n$ is calculable, but it is also possible to analytically 
continue it to non integer $n$. The method presented in \cite{solo} differs subtly from the above: instead of continuing analytically in $n$ 
the partition function $Z_n$, they first analytically continue the 
cover manifold assuming the corresponding geometry can be defined for an arbitrary and small angular deficit in the limit $n \rightarrow 1$ 
and then calculate the corresponding $Z_n$ \cite{solo,solo1}. 
Taking into account this is not possible in general, then, the identification of
the logarithmic coefficient with linear combinations of the stress tensor anomalies is not fully justified \cite{Myers2,Myers4,Schwimmer,Soloreview}. 
In the case of spherical sets, the objections to the applicability of the replication method disappear since there is an extra 
rotational symmetry in the transverse space about the entangling surface. In fact, the result for the sphere in four dimensions 
can be extended to any dimensions and any field theory \cite{cashuemye}. These results complete a solid 
proof for the term proportional to the type $a$ anomaly in (\ref{general}). 

On the other hand, for the cylinder, the situation is more obscure. Very recently, it was found in 
\cite{Myers2} for holographic models, that in four dimensions the logarithmic coefficient is proportional to the anomaly as expected, but that 
the same is not true for higher dimensions, where
new conformally invariant terms could be added to (\ref{general}). 
In this scenario, we conclude  our results give support to the validity of the Solodukhin's calculation for the cylinder in four dimensions.

\section*{Acknowledgments}
I thank H.Casini for useful discussions and comments on the manuscript.
I also acknowledge support from the EPLANET programme grant and Dr. G. Mussardo for his hospitality at SISSA where the initial stages of this work were done.
This work was partially supported by CONICET, ANPCyT and Universidad Nacional de Cuyo, Argentina.


\begin{thebibliography}{99}


\bibitem{ryu_taka}

  S.~Ryu and T.~Takayanagi,
  ``Holographic derivation of entanglement entropy from AdS/CFT,''
  Phys.\ Rev.\ Lett.\  {\bf 96}, 181602 (2006)
  [arXiv:hep-th/0603001].


\bibitem{ryu_taka2}   S.~Ryu and T.~Takayanagi,
  ``Aspects of holographic entanglement entropy,''
  JHEP {\bf 0608}, 045 (2006)
  [arXiv:hep-th/0605073].


\bibitem{solo}   S.~N.~Solodukhin,
  ``Entanglement entropy, conformal invariance and extrinsic geometry,''
  Phys.\ Lett.\  B {\bf 665}, 305 (2008)
  [arXiv:0802.3117 [hep-th]].


\bibitem{num}  R.~Lohmayer, H.~Neuberger, A.~Schwimmer and S.~Theisen,
  ``Numerical determination of entanglement entropy for a sphere,''
  Phys.\ Lett.\  B {\bf 685}, 222 (2010)
  [arXiv:0911.4283 [hep-lat]].

\bibitem{dowker} 
 
 J.~S.~Dowker,
  ``Entanglement entropy for even spheres,''
  arXiv:1009.3854 [hep-th],
   J.~S.~Dowker,
  ``Entanglement entropy for odd spheres,''
  arXiv:1012.1548 [hep-th].

\bibitem{cashue}  H.~Casini and M.~Huerta,
  ``Entanglement entropy for the n-sphere,''
  Phys.\ Lett.\  B {\bf 694}, 167 (2010)
  [arXiv:1007.1813 [hep-th]].

\bibitem{solo2} S.~N.~Solodukhin,
  ``Entanglement entropy of round spheres,''
  Phys.\ Lett.\  B {\bf 693}, 605 (2010)
  [arXiv:1008.4314 [hep-th]].

 \bibitem{dowker1} 
J.~S.~Dowker, ``Hyperspherical entanglement entropy,''
  J.\ Phys.\ A  {\bf 43}, 445402 (2010)
  [arXiv:1007.3865 [hep-th]].

\bibitem{cashuemye}  H.~Casini, M.~Huerta and R.~C.~Myers,
  ``Towards a derivation of holographic entanglement entropy,''
  JHEP {\bf 1105}, 036 (2011)
  [arXiv:1102.0440 [hep-th]].

\bibitem{Myers2} L.~Y.~Hung, R.~C.~Myers and M.~Smolkin,
  ``On Holographic Entanglement Entropy and Higher Curvature Gravity,''
  JHEP {\bf 1104}, 025 (2011)
  [arXiv:1101.5813 [hep-th]].


\bibitem{review}
 H.~Casini and M.~Huerta,
  ``Entanglement entropy in free quantum field theory,''
  J.\ Phys.\ A  {\bf 42}, 504007 (2009)
  [arXiv:0905.2562 [hep-th]].


\bibitem{sred} M.~Srednicki,
 ``Entropy and area,''
 Phys.\ Rev.\ Lett.\  {\bf 71}, 666 (1993)
 [arXiv:hep-th/9303048]. 

\bibitem{peschel}
I.~Peschel, J.\ Phys.\ A: Math.\ Gen. {\bf 36}, L205 (2003) [arXiv:cond-mat/0212631]. 

\bibitem{wilczek} M.~P.~Hertzberg and F.~Wilczek,
  ``Some Calculable Contributions to Entanglement Entropy,''
  Phys.\ Rev.\ Lett.\  {\bf 106}, 050404 (2011)
  [arXiv:1007.0993 [hep-th]].


\bibitem{Myers3}
   L.~Y.~Hung, R.~C.~Myers and M.~Smolkin,
  ``Some Calculable Contributions to Holographic Entanglement Entropy,''
  JHEP {\bf 1108}, 039 (2011)
  [arXiv:1105.6055 [hep-th]].

\bibitem{kabat}
  D.~N.~Kabat,
  ``Black hole entropy and entropy of entanglement,''
  Nucl.\ Phys.\  B {\bf 453}, 281 (1995)
  [arXiv:hep-th/9503016].

\bibitem{hlw}
 See for example, C.~Holzhey, F.~Larsen and F.~Wilczek,
 ``Geometric and renormalized entropy in conformal field theory,''
 Nucl.\ Phys.\ B {\bf 424}, 443 (1994)
 [arXiv:hep-th/9403108].

\bibitem{solo1}  S.~N.~ Solodukhin, Phys. \ Rev. \ D {\bf 51} (1995) 609, [arXiv:hep-th/9407001];
D.~V. ~Fursaev, S.~N. ~Solodukhin, Phys. Lett. B {\bf 365} (1996) 51, [arXiv:hep-th/
9412020];
D.~V. ~Fursaev, S.~N. ~Solodukhin, Phys.\ Rev.\ D {\bf 52} (1995) 2133, [arXiv:hep-th/
9501127]; R.~B.~ Mann, S.~N.~ Solodukhin, Nucl.\ Phys.\ B\ {\bf 523} (1998) 293, [arXiv:hep-th/
9709064];
S.~N.~ Solodukhin, Phys. \ Rev. \ D\ {\bf 57} (1998) 2410, arXiv:hep-th/9701106.
\bibitem{Myers4}
  R.~C.~Myers and A.~Sinha,
  ``Holographic c-theorems in arbitrary dimensions,''
  JHEP {\bf 1101}, 125 (2011)
  [arXiv:1011.5819 [hep-th]].
\bibitem{Schwimmer}
  A.~Schwimmer and S.~Theisen,
  ``Entanglement Entropy, Trace Anomalies and Holography,''
  Nucl.\ Phys.\  B {\bf 801}, 1 (2008)
  [arXiv:0802.1017 [hep-th]].

\bibitem{Soloreview}
  S.~N.~Solodukhin,
  ``Entanglement entropy of black holes,''
  arXiv:1104.3712 [hep-th].


\end{thebibliography}
\end{document}